\documentclass[aps,final,notitlepage,oneside,twocolumn,nobibnotes,nofootinbib,
superscriptaddress,noshowpacs,centertags]{revtex4-1}

\usepackage[english]{babel}
\usepackage{graphicx}
\usepackage{latexsym}
\usepackage{amssymb}
\usepackage{amsmath}
\usepackage{float}

\begin{document}

\title{Mergers of primordial black holes in extreme clusters and the $H_0$ tension}
\author{Yury Eroshenko}\thanks{e-mail: eroshenko@inr.ac.ru}
\affiliation{Institute for Nuclear Research of the Russian Academy of Sciences, Moscow, 117312 Russia}

\date{\today}

\begin{abstract}
We consider a cosmological model with dark matter in the form of $\sim10^{-12}M_\odot$ primordial black holes in dense weakly relativistic clusters with masses $18-560M_\odot$. It is shown that during the multiple collisions of the black holes the $\sim10$\% of the initial cluster mass can be transformed into gravitational waves in the time interval from recombination to the redshifts $z\geq 10$. At the recombination epoch, the density of matter was larger by $\sim10$\% and, accordingly, the universe expansion rate was higher. This leads to a shortening of the sound horizon scale, as is necessary to solve the ``$H_0$ tension'' problem.
\end{abstract}

\maketitle


\section{Introduction}

In recent years, the accuracy of cosmological measurements has significantly increased both in the study of relic radiation and in local measurements. This makes it possible to test the predictions of various cosmological inflation theories, for example, to measure small deviations of the cosmological perturbation spectrum from the Harrison-Zeldovich flat spectrum. However, a number of unresolved points remain. Apart from global questions about the nature of dark matter and dark energy, the structure of the universe as a whole, etc., there are some unexplained data (supermassive black holes at high redshifts, the EDGES absorption, etc.), as well as inconsistencies between different sets of measurements. 

One such recent problem is the ``$H_0$ tension'' or ``Hubble Tension'' problem \cite{BerDolTka15,Valetal20}. The nowadays Hubble constant as measured from the early universe by the Planck telescope  $H_0=67.37\pm0.54$~km/s/Mpc \cite{Aghetal20} is less than the Hubble constant $H_0=73.52\pm1.62$~km/s/Mpc \cite{Rieetal18} as measured in the local universe by Cepheids and Type Ia supernovae. Attempts were put to find an astrophysical solution without involving new physical phenomena. For example, in \cite{Yeretal20}, the problem is explained by the contribution of intergalactic dust to the measured cosmic microwave radiation. But many hypotheses go beyond the usual cosmological models. For example, the decays of dark matter particles \cite{BerDolTka15,ChuGorTka16,HryJod20,PanKarDas20}, the reannihilation of self-interacting dark matter \cite{Binetal18}, the generation of dark radiation, a certain type of dark energy evolution, etc. were considered (see the comprehensive review of solutions in \cite{Valetal21}).

One of the suggested approaches to achieve the agreement is to shift the sound horizon scale at the time of recombination, see for example, \cite{SetTod21}. Indeed, the first acoustic peak is on the angular scale $\theta_s=r_s/D$, where $D=c\int_{0}^{z_r}dz/H (z)$, $r_s=\int_{z_r}^{\infty}c_s(z) dz/H(z)$ is the sound horizon radius, $z_r$ is the recombination redshift, $c_s$ is the sound speed. The scale $\theta_s$ is fixed by the observational data. Suppose $D$ depends only on the Hubble constant in the current epoch. In this case, by measuring the scale of the first acoustic peak formed in the early universe, one can find the Hubble constant at the present time. But the value $H_0$ obtained in this way is less than it was measured in the local universe at the present time. It was suggested to eliminate the contradiction with local $H_0$ measurements by shortening the sound horizon $r_s$. Namely, it is assumed that the Hubble constant before recombination (in the integrand for $r_s$) was greater than in the standard picture. Therefore, the $r_s$ was smaller.

Therefore the ``$H_0$ tension'' problem could be explained by the non-relativistic matter mass transfer to the relativistic component (dark radiation) \cite{SetTod21}. Between the epoch of the mass transfer and the present time, this dark radiation has experienced a cosmological redshift and does not make a noticeable contribution to the modern Hubble constant. It is necessary that dark radiation appeared in the universe after the primordial nucleosynthesis, since it would give additional effective relativistic degrees of freedom, changing the yield of chemical elements.  See, however, the discussion in \cite{JedPogZha20}, where the difficulties of such models are noted. In addition, the difficulties with a change to late time physics as a solution were described in \cite{Efs21}.

In this paper, we also propose a model with the matter transfer into radiation. We consider the generation of gravitational waves when primordial black holes (PBHs) collide in dense clusters, assuming that all the PBHs make up the dark matter in the universe. When two PBHs merge, about 10\% of their mass is transformed into gravitational waves \cite{Hoo20,HeaLouZlo14}, and several consecutive mergers are possible in the PBHs clusters. If this energy transfer occurred after recombination, the density of dark matter in a comoving volume has decreased since recombination. The gravitational radiation has experienced a cosmological redshift to date and does not make a noticeable contribution to the current universe density. At the time of recombination, the density of dark matter in the comoving volume was $\sim10$\% higher, which increases the rate of cosmological expansion in that epoch compared to models without energy transfer. We show that this model is able to reduce the Hubble constant by $\sim5$\% and remove the ``$H_0$ tension'' problem. Note, that the value $\sim5$\% is requires for the intervals obtained in the works \cite{Aghetal20} and \cite{Rieetal18} began to overlap, although a smaller amount is sufficient to mitigate the discrepancies. The processes of PBHs formation and mergers lead to the emission of gravitational waves and the stochastic gravitational-wave background filling (see i.g. \cite{BugKli11,CleGar17,ManBirCho16,Wangetal18}).  The emission of the gravitational waves during the PBHs merges in pairs in the context of ``$H_0$ tension'' problem was discussed in \cite{RaiVasVee17}, where it was concluded that this mechanism cannot fully explain ``$H_0$ tension''. In our model, the PBHs merge not in isolated pairs, but in clusters, so the PBHs mergers rate and correspondingly the rate of the mass transfer into gravitational waves can be significantly higher than it was obtained in \cite{RaiVasVee17}.

The main ingredients of this model are PBHs and their clusters. The possibility of the PBH formation in the Universe was discovered in \cite{ZelNov66,Haw71}, and later many possible formation models were proposed, see the reviews \cite{Dol18,Beletal19,Caretal20}. There are many cosmological and astrophysical constraints on the number of PBHs in the universe in different mass intervals \cite{Caretal20}. However, the possibility is still allowed that the PBHs constitute all the dark matter \cite{CarKuh20,GreKav20}. In particular, the mass window  $10^{20}-10^{24}$~g still remains open. The emission of gravitational waves at the PBH mergers has already been discussed in a number of papers in various aspects. For example, in \cite{Hoo20}, the generation of the gravitational waves background by the low-mass PBHs before their quantum evaporation was considered. 

PBHs can be initially formed in the universe not uniformly, but in clusters \cite{RubSakKhl01,KhlRubSak02,Beletal19,Beretal20}. One of the well-developed models for the PBH clusters is the domain wall fragmentation model \cite{RubSakKhl01,KhlRubSak02}. In this model, both the PBHs themselves and their clusters can be formed with different masses and scales, depending on the parameters of the theory. In this regard, the masses of the PBHs and the parameters of the clusters at the current stage can be considered as free parameters, because the theory and observations does not fix them. 

In this paper, without specifying a particular mechanism of PBHs and clusters formation, we assume that the dark matter in the universe consists of PBHs with masses $10^{20}-10^{24}$~g. Another our assumption is that PBHs are in very compact clusters, which are already at the weakly relativistic state, and their further evolution leads to the evaporation of PBHs from the clusters and the contraction of the remaining core. We may call such systems by ``extreme clusters''. There are no definite predictions for the their formation. The presence of the extreme clusters in the Universe with the necessary parameters is the main assumption of this work. In this paper, we present an example of parameters that solve the $H_0$ tension by the mass transfer from the PBHs into the gravitational waves. As we will show, this model is self-consistent in the general cosmological picture.


\section{Friedman equations with mass flow}

The Friedman equations for a flat cosmological model are written as
\begin{equation}
H^2=\left(\frac{1}{a}\frac{da}{dt}\right)^2=\frac{8\pi 
G}{3}\rho,
\label{h2}
\end{equation}
\begin{equation}
\frac{d^2a}{dt^2}+\frac{4\pi G}{3c^2}a(\rho c^2+3p)=0.
\label{doth}
\end{equation}
The consequence of these two equations is 
\begin{equation}
\frac{d\rho}{dt}=-3H(\rho+p/c^2).
\label{term1}
\end{equation}
At the same time
\begin{equation}
\rho=\rho_m+\rho_r+\rho_\Lambda, \quad p/c^2=\frac{\rho_r}{3}-\rho_\Lambda,
\end{equation}
and $\rho_\Lambda=const$.

As it follows from the energy conservation, in the elementary processes of matter transformation into radiation, the sum $\rho_m+\rho_r$ does not change, so the appearance of relativistic particles does not affect the Hubble constant directly through the equation (\ref{h2}). But the change in the pressure affects the evolution of $\rho_m+\rho_r$ over time through the equation (\ref{doth}), and thus there is an indirect influence on the Hubble constant. Note also that the curvature of space cannot change due to internal transformations of energy and pressure. In the presence of the energy redistribution between matter and radiation, instead of (\ref{term1}) one s the following equations, which are actually a consequence of the second law of thermodynamics, 
\begin{equation}
\frac{d\rho_m}{dt}=-3H\rho_m-Q,
\label{term21}
\end{equation}
\begin{equation}
\frac{d\rho_r}{dt}=-4H\rho_r+Q,
\label{term22}
\end{equation}
where $Q(t)$ is the density flow between matter and radiation. Together with the equation (\ref{h2}), these two equations form a complete system describing the problem (of course, if the function $Q(t)$ is somehow known). Next, instead of $Q(t)$, it will be convenient to consider the mass flow in the comoving volume $q(t)=Q(t)[a(t)/a_0]^3=Q(t)/(1+z)^3$, where the index ``0'' denotes the values at the present time $t_0$.

For $q=0$, let the density of matter and radiation currently be $\rho_{m,0}$ and $\rho_{r,0}$, respectively. By matter we mean the baryons and non-relativistic dark matter, and by radiation we mean all the relativistic components. In the presence of a mass flow, these quantities are $\rho_{m,0}-\Delta$ and $\rho_{r,0}+\Delta$, respectively, where $\Delta$ is some arbitrary but small ($\Delta\ll \rho_{m,0}$) value, meaning that their sum is the same and is determined by the local Hubble constant. Then the solutions of the equations (\ref{term21}) and (\ref{term22}) have the form
\begin{equation}
\rho_m(t)=(1+z)^3\rho_{m,0}-(1+z)^3\Delta+(1+z)^3\int\limits_{t}^{t_0}q(t)dt
\label{sol1}
\end{equation}
and
\begin{equation}
\rho_r(t)=(1+z)^4\rho_{r,0}+(1+z)^4\Delta-(1+z)^4\int\limits_{t}^{t_0}\frac{q(t)dt}{1+z}.
\label{sol2}
\end{equation}

\begin{figure}
	\begin{center}
\includegraphics[angle=0,width=0.49\textwidth]{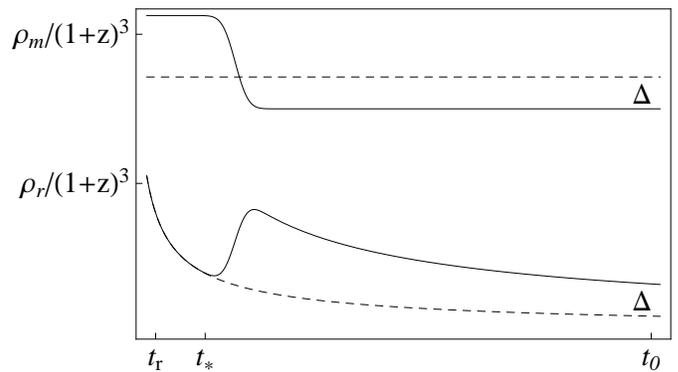}
	\end{center}
\caption{An example of the matter and radiation comoving density evolution according to (\ref{sol1}) and (\ref{sol2}) with the mass-to-radiation transformation begun at the moment $t_*$. The dashed line show the densities without the transformation (if $q=0$). Units on the axes are conditional.}
\label{grex}
\end{figure}

Let $t_*$ be the moment when the density transfer starts, before which $q=0$, then
\begin{equation}
-\Delta+\int\limits_{t_*}^{t_0}\frac{q(t)dt}{1+z}=0.
\label{dint}
\end{equation}
And for $t<t_*$ one have
\begin{equation}
\rho_m(t)=(1+z)^3\rho_{m,0}+(1+z)^3\int\limits_{t}^{t_0}\frac{z}{1+z}q(t)dt.
\label{sol1tless}
\end{equation}
Thus, if the energy transfer was completed at a large $z$, then the increase in the density of matter $\rho_m$ in the comoving volume in the early epochs can be estimated as
\begin{equation}
\rho_m(t)\simeq(1+z)^3\rho_{m,0}+(1+z)^3\int\limits_{t}^{t_0}q(t)dt,
\label{sol1itog}
\end{equation}
since $z/(1+z)\simeq 1$ at $z\gg1$. We assume that the transfer was completed at $z\geq10$, that is, before the time $t\simeq t_{10}\simeq1.5\times10^{16}$~sec.
An example of the matter and radiation comoving density evolution according to (\ref{sol1}) and (\ref{sol2}) is shown at the Fig.~\ref{grex}.

Since in the ``$H_0$ tension'' problem the difference is only a few percent, we will search for a solution of the Friedman equations by the method of successive approximations, taking as the zero approximation the case $q=0$, in which 
\begin{equation}
H^2=\frac{8\pi G}{3}\bar\rho,
\end{equation}
where $\bar\rho$ is the sum of the densities of matter, radiation, and the lambda-term in the case $q=0$. If in the real situation (for $q\neq0$) the density is $\bar\rho(t)+e(t)$, then from the perturbed Friedman equation 
\begin{equation}
(H+\delta H)^2=\frac{8\pi G}{3}\left[\bar\rho(t)+e(t)\right]
\end{equation}
we get the relative correction to the Hubble constant in the early times $t<t_*$
\begin{equation}
\frac{\delta H}{H}\simeq\frac{1}{2}\frac{e(t)}{\bar\rho(t)}\simeq\frac{1}{2\rho_{c,0}\Omega_m}\int\limits_{t_*}^{t_0}q(t)dt,
\label{itogh}
\end{equation}
where $\rho_{c,0}$ is the current critical density, and the latter estimate is applicable at $t_{\rm eq}<t<t_*$, including the $t\sim t_r$ region, where $t_{\rm eq}$ is the moment of equality. In this method of successive approximations, the dependence of $t(z)$ is described by the usual expression known from the zero approximation (when $q=0$) 
\begin{equation}
H^2=H_0^2\left[\Omega_r(1+z)^4+\Omega_m(1+z)^3+\Omega_\Lambda\right],
\label{h2new}
\end{equation}
where the cosmological density parameters are taken at the present time. 

To solve the ``$H_0$ tension'' problem, it is necessary that the value (\ref{itogh}) at the epoch $t\sim t_r$ was about 5\%. Thus, it follows from (\ref{itogh}) that at $t<t_{10}$ about 10\% of the nonrelativistic matter should be transformed into radiation, so that at $t\sim t_r$ the Hubble constant was $\sim5$~\% greater than in the $q=0$ case. This radiation will not contribute noticeably to the current density due to the cosmological redshift, and the current value of $H_0$ will not change under the accepted assumptions. Since, as we assume, the mass transfer started quite late, the additional radiation did not contribute to the number of relativistic degrees of freedom during the primordial nucleosynthesis, which avoids a contradiction with the theory of primordial nucleosynthesis. In the next chapters, we will show that the described situation could be realized in the presence of dense PBH clusters. PBHs mergers in clusters with the density transfer from the non-relativistic component into gravitational waves, providing the necessary shift in the Hubble constant at early epochs.


\section{Dynamical evolution of black hole clusters}

Let there is a cluster of $N$ PBHs with masses $m$, total mass $M=Nm$, and radius $R$. These values change over time during the dynamical evolution of the cluster, and their initial values at the time of cluster formation will be marked with the index ``i''. The moment of the cluster formation $t_i$ probably corresponds to the moment of its entry under the cosmological horizon. In this case, for the parameters used below, $t_i\ll t_r$. As we get from the equations of cluster evolution, the cluster evolves slowly at first, but at the very last stage the evolution accelerates very much, and finally the cluster evaporates or collapses.

The dynamical evolution of the PBHs cluster occurs due to two-body relaxation and due to the capture of PBHs in short-lived close binary systems, which quickly emit kinetic energy into gravitational waves and merge \cite{QuiSha87}. The first stage of evolution, when mergers are rare, is called non-dissipative. At this stage, the fastest PBHs fly out of the cluster, carrying away energy (dynamic evaporation). The remaining cluster mass $M(t)$ decreases and the cluster is compressed with an increase in the velocity dispersion $v\simeq[GM/(2R)]^{1/2}$. The loss in the tail of the velocity distribution is constantly restored due to dynamic relaxation, and all the time there are fast PBHs that leave the cluster. When the remaining part of the cluster shrinks sufficiently, dissipative processes begin in it, associated with the emission of gravitational waves during close PBHs flyby and mergers. During the transition to the dissipative stage, the evolution of the cluster is greatly accelerated, and as a result, the cluster completely evaporates or collapses. Gravitational collapse is possible when the velocity dispersion in the cluster approaches the speed of light. The collapse occurs in an avalanche-like manner due to the presence of overlapping PBH orbits in the cluster.

\begin{figure}
	\begin{center}
\includegraphics[angle=0,width=0.49\textwidth]{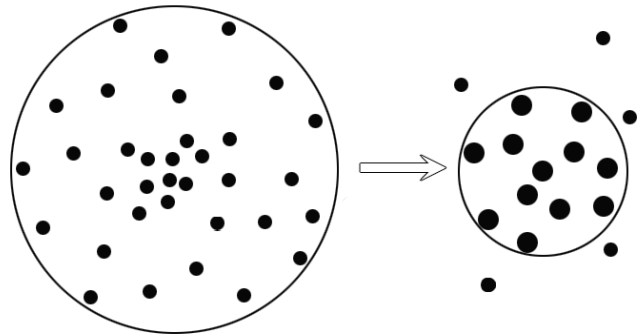}
	\end{center}
\caption{Schematic view of the cluster evolution. PBHs merge into the larger PBHs and some of them fly away from the cluster (dynamical evaporation). The remaining cluster gradually contracts until the final disappear or collapse.}
\label{grcluster}
\end{figure}

In the process of the merger, the PBHs can radiate about 10\% of its mass into gravitational waves. This is exactly the process that in our model is responsible for mass transfer into the relativistic component (gravitational waves). The individual PBH masses grow $m=m(t)$. We don't consider the PBH mass distribution, i.e. we suppose that all the objects have the same mass.  The schematic view of the cluster evolution is shown at Fig.~\ref{grcluster}. Let us introduce the variable $x=v^2/c^2$. The $x$ is simply $1/4$ the ratio of the cluster gravitational radius $2GM/c^2$ to its radius $R$. Approximately, it is assumed that the cluster has a uniform density. In this approximation, the equations describing the cluster evolution were obtained in \cite{QuiSha87}. Using the variable $x$, these equations are written as
\begin{equation}
\frac{1}{x}\frac{dx}{dt}=\frac{(\alpha_2-\alpha_1)}{t_r}+\frac{7}{5t_{\rm cap}},
\label{gom1} 
\end{equation}
\begin{equation}
\frac{1}{M}\frac{dM}{dt}=-\frac{\alpha_2}{t_r},
\label{gom2} 
\end{equation}
\begin{equation}
\frac{1}{m}\frac{dm}{dt}=\frac{1}{t_{\rm cap}},
\label{gom3} 
\end{equation}
where $t_{\rm cap}^{-1}=\sigma_{\rm cap} v n/\sqrt{2}$, $n$ is the number density of the PBHs in cluster, 
\begin{equation}
\sigma_{\rm cap}\approx\frac{6\pi G^2m^2}{c^{10/7}v^{18/7}},  
\label{sigm1m2}
\end{equation}
and the two-body relaxation time is
\begin{equation}
\label{tr}
t_{\rm r}=\left(\frac{2}{3}\right)^{1/2}
\frac{v^3}{4\pi G^2m^2n\Lambda},
\end{equation}
where $\Lambda=\ln(0.4 N)\simeq7$. The presence of the Coulomb logarithm $\Lambda$ emphasizes that close and distant approaches are equally important for the two-body dynamic evolution of the cluster. Next, we will assume approximately that $\Lambda=const$.
According to \cite{QuiSha87}, the constants $\alpha_1=8.72\times10^{-4}$, and $\alpha_2=1.24\times10^{-3}$ were obtained in the Fokker-Planck cluster model (see \cite{QuiSha87} for  a more detailed description and references).

In \cite{DokEro12}, the exact solution of the system (\ref{gom1})-(\ref{gom3}) was found via the variable $x$ in the approximation $\Lambda=const$. By introducing the notations
\begin{equation}
x_{\rm dis}\equiv\left[\frac{10\Lambda(\alpha_2-\alpha_1)}{(7\sqrt{3})}\right]^{7/5}
\simeq 1.8\times10^{-4},
\end{equation}
\begin{equation}
y\equiv \left(\frac{x}{x_{\rm dis}}\right)^{5/7},
\end{equation}
and
\begin{equation}
\varkappa\equiv\frac{7\alpha_2}{5(\alpha_2-\alpha_1)}\simeq4.7,
\end{equation}
this solution can be written as
\begin{equation}
m(y)=m_i\frac{1+y}{1+y_i},
\label{mtmi}
\end{equation}
\begin{equation}
M(y)=M_i\left[\frac{(1+y)y_i}{y(1+y_i)}\right]^\varkappa. 
\label{mbigt}
\end{equation}

The beginning of the dissipative stage of evolution corresponds to $x\sim x_ {\rm dis}$, i.e., $v/c\simeq0.013$, $v\simeq4000$~km~s$^{-1}$ regardless of other cluster parameters.  The gravitational collapse of the cluster occurs at $x\simeq x_f\simeq0.1$ \cite{QuiSha87}. The collapse corresponds to $y\sim y_{\rm col}\simeq 91$.  According to (\ref{mtmi}) and (\ref{mbigt}), by the time of collapse, the masses of the black holes could have grown to $\sim46m_i$, and the collapsing remainder of the cluster would have a mass of $0.04M_i$. However, with our choice of parameters specified below, the cluster does not reach the collapse stage, but completely dynamically evaporates due to the flow of the PBHs from it. However, before the evaporation of the cluster, the PBHs in the cluster managed to merge several times with each other and, as we will show soon, the total mass transformed into gravitational waves can reach $\sim10$\% of the total initial cluster mass  $M_i$, which is enough to solve the ``$H_0$ tension'' problem.


\section{Transformation of mass into gravitational waves in a dense cluster}

As a fairly conservative assumption to avoid contradictions with the observational data, we assume that the density transfer took place in the time interval from $t_r$ to $t_{10}$. In theory, the mass transfer could have occurred earlier, since the clusters have a very early origin, but early transfer at $t<t_r$ requires denser clusters, which, moreover, collapse early.  We will based on a more conservative version with the mass transfer at $t_r\leq t\leq t_{10}$, which does not require very dense clusters and fine-tuning of parameters. Our goal is to show that at least in this case, it is possible to obtain reasonable cluster parameters and to provide a solution to the ``$H_0$ tension'' problem.

The number of the merging BHs increases according to the equation
\begin{equation}
\frac{d N_c}{dt}=\frac{N}{t_{\rm cap}}.
\end{equation}
The merging mass is $dM_c=mdN_c$, so when the parameter $y$ is changed from $y_i$ to $y$, the merged total mass is
\begin{equation}
\frac{M_c}{M_i}=\frac{y_i^\varkappa}{(1+y_i)^\varkappa}\int\limits_{y_i}^y\frac{(1+y)^{\varkappa-1}dy}{y^{\varkappa}}.
\label{mcbig}
\end{equation}
The value (\ref{mcbig}) can exceed one, since the mass of each PBH could participate in several consecutive mergers. This statement can be illustrated by the simple toy example. Let's we have four PBH with masses $m$ and the total mass $M_i=4m$. Let the PBHs merged in pairs. After the first merger, we have two PBH with masses $2m$, and the merged mass is $4\times m=4m$. Let the resulting two PBH merge into a single PBH with mass $4m$. In this second merge, the mass $2\times 2m=4m$ is merged. During the two cycles of mergers, the total mass $4m+4m=8m$ was merged. Thus, the total merged mass is $M_c=8m=2M_i$. Without dynamical evaporation of the PBHs from a cluster, the maximum mass $M_c \sim M_i\log_2 N_i$ could merge in it, where $N_i$ is the initial number of PBHs in the cluster. In reality, the cluster loses PBHs due to dynamical evaporation, and the merged mass is much smaller.

\begin{figure}
	\begin{center}
\includegraphics[angle=0,width=0.49\textwidth]{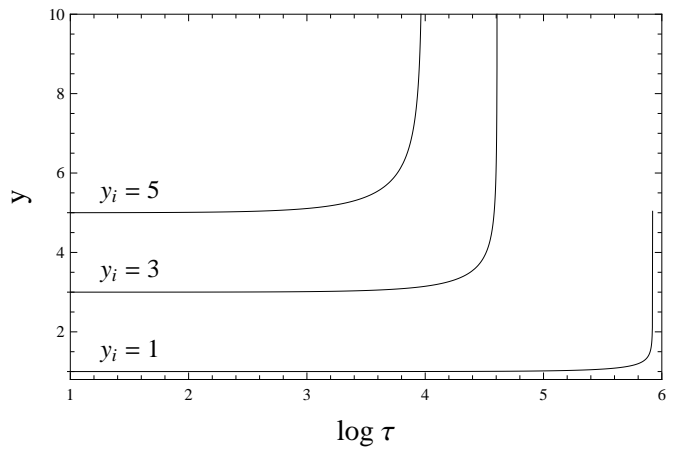}
	\end{center}
\caption{Solution $y(\tau)$ of the equation (\ref{yeq}) in dimensionless time $\tau$ for (from bottom to top) $y_i=$1, 3, 5.}
\label{grytau}
\end{figure}
Now we calculate the parameters of clusters that collapse or evaporate in the time interval from $t_r$ to $t_{10}$. As mentioned above, this gives a fairly conservative estimate. To trace the dynamics of the cluster over time, one needs to solve numerically  the equation (\ref{gom1}), which we will rewrite as
\begin{equation}
\frac{1}{y}\frac{dy}{dt}=\frac{5(\alpha_2-\alpha_1)}{7t_r}+\frac{1}{t_{\rm cap}}.
\label{yeq}
\end{equation}
To solve this equation, it is convenient to introduce a dimensionless variable $\tau=tc^3/(Gm_iN_i^2)$. With this variable we have 
\begin{eqnarray}
\frac{dy}{d\tau}&=&\frac{f_1(y)}{f_2^2(y)}y^{31/10}\left[2^{5/2}3^{3/2}5(\alpha_2-\alpha_1)x_{\rm dis}^{3/2}/7+\right.
\nonumber
\\
&+&\left.2^{3/2}3^2x_{\rm dis}^{31/14}y\right],
\label{yeqdl}
\end{eqnarray}
where $f_1(y)=(1+y)/(1+y_i)$, and $f_2(y)=[((1+y)y_i)/(y(1+y_i))]^\varkappa$. The value in square brackets in (\ref{yeqdl}) in the numeric representation has the form $\simeq1.7\times10^{-8}[1+7y]$, i.e. for extreme clusters, the second term prevails. An approximate solution can be obtained by substituting $y=y_i$ in $f_1(y)$ and $f_2(y)$ and neglecting the first term in square brackets in (\ref{yeqdl}), 
\begin{equation}
y\simeq \frac{y_i}{\left[1-4\times10^{-7}y_i^{31/10}\tau\right]^{10/31}}.
\label{rough1}
\end{equation}
From here we roughly get the time when the rapid growth of $y$ begins,
\begin{equation}
\tau_b\sim 2.5\times10^6y_i^{-31/10}.
\label{rough2}
\end{equation}

The exact numerical solution of (\ref{yeqdl}) shows that $y (\tau)$ remains near the initial value for a long time, experiencing a very slow growth, and then at the final stage, for just one order of $\tau$, a sharp rise begins to formally infinite values in a finite time $\tau$ (the $y$ of course cannot be larger than $1/x_{\rm dis}^{5/7}$). This is due to the structure of the equation (\ref{yeqdl}). Examples of the evolution $y(\tau)$ are shown in Fig.~\ref{grytau}. This sharp increase occurs near the values $\tau\simeq\tau_b\simeq 8.4\times10^5$, $4.1\times10^4$, and $9.5\times10^3$ for $y_i=$1, 3, and 5, respectively.  By an order of magnitude, these values are near the solution (\ref{rough2}). The $\tau_b$ corresponds to the time $t_b=\tau_bGm_iN_i^2/c^3$.

Let's choose $y_i=5$ as an example. These are fairly dense weakly relativistic clusters with $v/c\simeq0.04$, so we call them ``extreme clusters''. Despite their extreme nature, they evolve quite slowly and completely evaporate without experiencing collapse. From the above-discussed requirement $t_r<t_b<t_{10}$, we obtain the following conditions for the initial masses of clusters
\begin{equation}
18\left(\frac{m_i}{10^{-12}M_\odot}\right)^{1/2}\leq\frac{M_i}{M_\odot}\leq 560\left(\frac{m_i}{10^{-12}M_\odot}\right)^{1/2}.
\label{miusl}
\end{equation}
From the cluster evaporation condition $M(y)=m(y)$, one get the parameter $y$ at the moment of evaporation $y_{\rm ev}=27$ and 41 for the left and right sides of the inequality (\ref{miusl}), respectively. For these values, from (\ref{mcbig}) we get $M_c(y=y_{\rm ev}, M_i=18M_\odot)=1.01M_i$, $M_c(y=y_{\rm ev}, M_i=560M_\odot)=1.21M_i$. In these cases, $\sim10$~\% of the initial mass $M_i$ of the cluster is transformed into gravitational waves, since such a part of the mass passes into gravitational waves at each merger of the PBHs, regardless of which generation they belong to, $m_i$, $2m_i$,.... Note that the collapse would correspond to $y\simeq91$, so, as mentioned above, the clusters with $y_i=5$ do not collapse, but completely evaporate, because $y_{\rm ev}<y_{\rm col}\simeq91$ for them.

If some of the clusters were even more relativistic ($y_i>5$) at the time of their formation, then they could quickly collapse into the single black holes. If only $\sim10^{-4}-10^{-3}$ fraction  \cite{Sasetal16} of the clusters experienced such collapses, then the resulting massive PBHs could explain the gravitational wave signals of LIGO/Virgo. The corresponding masses given by the constraint (\ref{miusl}) just fall into the mass range required for LIGO/Virgo events, but the left part of this interval is more preferable. In reality, the masses of PBHs formed during collapses should be somewhat smaller due to the dynamic evolution of clusters from the moment of formation to collapse.

Like the models with single large PBHs, the clusters must be randomly distributed, according to the Poissonian law. This follows from the generation of the seed perturbations at the inflation stage. In the different Hubble volumes, the perturbations are independent, so the Poissonian distribution is realized. In this regard, there is no difference in the statistics of the PBH pairs formed as a result of  the early clusters collapses and the single large  PBHs. Thus, the statistics of LIGO/Virgo signals lead to the same proportion $\sim10^{-4}-10^{-3}$ of the stellar mass PBHs in the composition of all dark matter, as in the case of non-clustered PBHs \cite{Sasetal16}.

The distribution of clusters by parameters is even more natural than the formation of them with some fixed parameters. This is due to the fact that perturbations are generated from quantum fluctuations at the inflation stage with an approximately Gaussian distribution. The maximum of the distribution corresponds to the typical values, and the tail of the distribution corresponds to the clusters with the most extreme values. 
The presence of a distribution will cause some part of the clusters to evolve at earlier times, and the PBH mergers will give some contribution to the radiation. If we knew the distribution of clusters by parameters, we could calculate this contribution. At this stage, we can only state that this contribution is very small, because in earlier epochs, the average density of the PBH in the Universe was much less than the density of the relic radiation. Indeed, the density of the PBH varies as $(1+z)^{-3}$, and the radiation density as $(1+z)^{-4}$, so their ratio from the moment of equality changes as $(1+z)^{-1}$. At the time of primordial nucleosynthesis, this ratio was extremely small, so the emission of the gravitational waves could not give noticeable contribution to the relativistic degrees of freedom.


\section{Conclusion}

In this paper, we assumed that in the early universe dense PBH clusters appeared by some mechanism with the masses of individual PBHs $10^{20}-10^{24}$~g, and the velocity dispersion approaching weakly relativistic values. If the initial masses of these clusters are within the interval, $18-560M_\odot$, then the further dynamical evolution of these clusters occurs in such a way that about $\sim10$\% of their initial mass could be transformed into gravitational waves during the merger of the PBHs. This means that in the early universe, during the recombination epoch, the density of dark matter per unit of comoving volume was about $\sim10$\% greater than in the modern epoch. Accordingly, the Hubble constant was $\sim5$\% higher than it would have been without the process of mass transformation into gravitational waves. Increasing the Hubble constant in the recombination epoch $H(t_r)$ leads to the shortening of the first acoustic peak scale. Thus, from measurements in the early Universe, the nowadays Hubble constant $H_0$ would be less than from the measurements in the local universe.

It is not possible to trace the evolution of the PBHs cluster in detail at the level of the simple estimates, because in the course of evolution, due to the randomness of the merger process, the masses of PBHs do not grow simultaneously, and, as a result, the mass segregation develops. To study this process, numerical simulations are needed, similar to those performed for the case of stellar mass PBHs. The computational complexity of this case is the large number of PBHs in the cluster $N_i\sim10^{14}$. However, our simple estimates already show that the mechanism of mass transformation into gravitational waves can be effective and can explain the $H_0$ tension.

The presence of extreme PBH clusters in the Universe may have a number of other consequences, as well as differences from the models with unclustered PBHs. The accretion of baryonic gas onto the PBH clusters in the early epochs may have a different efficiency than if the PBHs were distributed uniformly. This may shift the accretion constraints on the PBHs. PBH clusters at the cosmological epoch $z>10$ could serve as seeds for the first stars, and perhaps even more massive black holes. The gas inhomogeneities caused by the clusters could create fluctuations in the 21~cm neutral hydrogen line. Some of the less dense clusters may have survived to the present time and their searches for can be carried out in the astronomical observations, i.g. through the mirolensing effect. Unlike a single large PBH, a cluster would be a non-compact lens with different light curve. However, the study of these and other possible extreme clusters manifestations is beyond the scope of the current work.

Summing up, the model considered in this paper can describe three cosmological phenomena at once:

\begin{itemize}

\item PBHs with masses $10^{20}-10^{24}$~g, those that have flown out of clusters during their evolution make up the bulk of dark matter. For this mass range, there are only weak model-dependent constraints on the PBHs, so the dark matter models with such PBHs are acceptable.

\item Approximately $\sim10^{-4}-10^{-3}$ fraction of the densest and early-collapsing clusters could have turned into PBHs with masses $\sim18-560M_\odot$ and be the black holes that are observed by LIGO/Virgo.

\item The emission of gravitational waves during the mergers of PBHs in the extreme clusters can explain the ``$H_0$ tension'' problem.

\end{itemize}


\section{Acknowledgements}
The author is grateful to V. K. Dubrovich for useful discussions of the ``$H_0$ tension'' problem,  and to S. G.~Rubin for the discussions about the clusters formation. The author is very grateful to the anonymous Referee for many valuable comments and suggestions,  which improved the paper significantly.

\end{document}